\documentclass[
aps,
pra,
twocolumn,
superscriptaddress,notitlepage]{revtex4-1}

\usepackage{amsmath}
\usepackage{amsfonts}
\usepackage{amssymb}
\usepackage{array}
\usepackage{floatflt}
\usepackage{graphicx}
\usepackage{color}
\usepackage{soul}

\begin{document}
\title{Global sensitivity analysis for optimization of the Trotter-Suzuki decomposition}

\author{Alexey N. Pyrkov}
\email{pyrkov@icp.ac.ru}
\affiliation{Institute of problems of chemical physics RAS, Acad. Semenov av. 1, Chernogolovka, Moscow region, Russia, 142432}

\author{Yurii Zotov}
\affiliation{Huawei Russian Research Institute, Huawei Technologies Co. Ltd., Moscow, Russia, 121614}

\author{Jiangyu Cui}
\affiliation{Central Research Institute, Huawei Technologies, Shenzhen 518129, China}

\author{Manhong Yung}
\affiliation{Data Center Technology Laboratory, Huawei Technologies Co Ltd, 115371 Shenzhen, Guangdong, China}


\begin{abstract}

The Trotter-Suzuki decomposition is one of the main approaches for realization of quantum simulations on digital quantum computers. Variance-based global sensitivity analysis (the Sobol method) is a wide used method which allows to decompose output variance of mathematical model into fractions allocated to different sources of uncertainty in inputs or sets of inputs of the model. Here we developed a method for application of the global sensitivity analysis to the optimization of Trotter-Suzuki decomposition. We show with a proof-of-concept example that this approach allows to reduce the number of exponentiations in the decomposition and provides a quantitative method for finding and truncation 'unimportant' terms in the system Hamiltonian.

\end{abstract}

\maketitle

\section{Introduction}

Nowadays complex mathematical and computational models are used in all areas of modern society and affected economy and technologies in many different ways \cite{der2019}. Contemporary mathematical models and development of computational devices have given impulse to fast digitalization of different areas from medicine to agriculture \cite{lecun15, ghahramani15, esteva19, carrio17}. However, growing of data and complexity of systems under consideration does not allow to process all that exactly and the ab-initio understanding of important features of the complex mathematical models looks like an unrealizable dream \cite{gorban2020,donoho2000}. Recently, a lot of methods were developed in order to extract the most important features of the complex models in such a manner that it is possible to make predictions about the outcomes on the basis of the chosen features only \cite{samet2006, lee1999, roweis2000, boehmke2019, antoulas2005}. One of such methods is the variance-based global sensitivity analysis (the Sobol method) \cite{sobol1993,sobol2001,saltelli2010,saltelli2008}. It is a wide used method which allows to decompose output variance of mathematical model into fractions allocated to different sources of uncertainty in inputs or sets of inputs of the model. This allows to simplify model under consideration: we can identify the inputs that have no effect on the output and remove redundant parts of the model from consideration.

On the other hand, recently the fast progress in experimental quantum computing and quantum algorithms was observed \cite{biamonte2017,carleo2019,jeswal2019,broughton2020,lazarev2020,cong2019,ibm,lu2019,nautrup2019,gao2018,melnikov2018,peruzzo2014} and now many researchers are trying to learn how quantum computing and the Big Data can benefit from each other. Since the idea of quantum computing appeared, one of the most promising applications of quantum computers has been the simulation of complex quantum systems \cite{feynman82, lloyd96, berry2007, malley16, guzik05, kassal2011, cody2012, babbush18}. Nowadays quantum simulations has become one of the most important applications for NISQ devices \cite{preskill18, zhang17, freitas18, poulin18, kivlichan19, lanyon11, langford17, childs18, gross17}. Quantum programmable devices with two-qubit gate fidelities more 98 percents were developed \cite{zajac18, huang19, barends14, wright19, gaebler16,ballance16,rol19,kjaergaard19} and this raises the exciting opportunities for solving problems that are beyond the reach of classical computation \cite{arute19,preskill18}. However, the error rates of currently available and near-term quantum machines are severely limited by the total number of gates
that can be reliably performed to simulate the dynamics of quantum systems. Usually, the total number of gates is defined by the Trotter-Suzuki (TS) decomposition \cite{trotter59, suzuki91, ruth83,childs2019}.  This decomposition allows to represent the evolutionary operator, in the first order expansion, as a product of matrix exponents of the system Hamiltonian terms. Furthermore, TS decomposition is a key routine in some methods of quantum optimal control \cite{Henneke2015}, in time evolution algorithms for matrix product states \cite{vidal2004,white2004,orus2008} and so on \cite{kolorenc2011}. Thus reaching the goal of practical quantum supremacy will require not only significant experimental advances, but also careful quantum algorithm design and implementation. Recently, some methods for optimization of TS decomposition were proposed. In particular, the schemes based on evolutionary algorithms \cite{jones2019}, random sampling \cite{campbell2019} and so on \cite{barthel2020} were considered. Nevertheless, all the protocols work with some restrictions and have some disadvantages for practical using on current quantum devices (for example, the protocol on the basis of random sampling \cite{campbell2019} has advantages only for the huge unpractical number of gates unachievable on current digital quantum machines). On the other hand, one of the simplest approaches that can be used on the current generation of quantum devices and as preprocessing for other more complex protocols is optimization of TS decomposition by removing unimportant gates from consideration with some increasing in error. However, in many cases it is not clear what terms of Hamiltonian can be removed from consideration when we apply the TS decomposition and what criteria we can use for that. Nowadays, the removing of the irrelevant terms is mostly empirical and deals only with the terms which are a few orders of magnitude smaller in terms of some matrix norm in comparison with others.

Here, we develop an approach based on the Global sensitivity analysis for optimization of the TS decomposition to remove the irrelevant terms. It is shown that the approach allows to find most of the terms that give small contribution in the whole problem variance providing a method for quantitative analysis of the irrelevant terms in the system Hamiltonian. We show that the removing these terms from consideration allows to reduce the number of exponentiations up to 25 percent with moderate increasing in the error of approximation.

\section{The Trotter-Suzuki decomposition}

Simulations of quantum dynamics of complex systems is the classically unsolvable problem due to the fact that the dimension of the problem grows exponentially with the system size and the matrix exponentiation is required. The Trotter-Suzuki decomposition theoretically allows one to model efficiently evolution of complex quantum systems on a digital quantum computer. This decomposition approximates the matrix exponentiation of the sum of non-commuting operators with the product of exponents, each of which can be represented through single-qubit and two-qubit gates, and thus allows to simulate the matrix exponentiation of quantum systems  with the desired accuracy.

The simplest first-order Trotter-Suzuki decomposition can be represented as
\begin{equation}\label{first_order}
e^{x(A+B)}\approx e^{xA}e^{xB}+ O(x^2),
\end{equation}
where $x$ - small parameter and $A$, $B$ - non-commuting operators $[A,B]\neq0$.

In order to obtain the second--order expansion, it is possible to represent both sides of the equation ~ (\ref{first_order}) in the following form:
\begin{multline}\label{second_order}
e^{x(A+B)}=I+x(A+B)+\frac{1}{2}x^2(A+B)^2+O(x^3)=\\
I+x(A+B)+\frac{1}{2}x^2\left(A^2+AB+BA+B^2\right)+O(x^3),\\
e^{xA}e^{xB}=\left(I+xA+\frac{1}{2}x^2A^2+O(x^3)\right)\times\\
\left(I+xB+\frac{1}{2}x^2B^2+O(x^3)\right)=\\
I+x(A+B)+\frac{1}{2}x^2\left(A^2+2AB+B^2\right)+O(x^3).
\end{multline}
Thus, we get the following expression
\begin{equation}\label{second_order2}
e^{xA}e^{xB}=e^{x(A+B)+\frac{1}{2}x^2[A,B]+O(x^3)},
\end{equation}
which corresponds to the well-known Baker-Hausdorff operator identity
\begin{equation}
e^{(A+B)}= e^{A}e^{B}e^{-\frac{1}{2}[A,B]},
\end{equation}
when $ A $ and $ B $ commute with their commutator.

Usually, using the Trotter-Suzuki decomposition, we split $x$ into $n$ parts and in this case the representation can be written as:
\begin{multline}
\left(e^{\frac{x}{n}A}e^{\frac{x}{n}B}\right)^n
=\left[
e^{\frac{x}{n}(A+B)+\frac{1}{2}\left(\frac{x}{n}\right)^2[A,B]+O\left(\left(\frac{x}{n}\right)^3\right)}
\right]^n \\
=e^{x(A+B)+\frac{1}{2}\frac{x^2}{n}[A,B]+O\left(\frac{x^3}{n^2}\right)},
\end{multline}
that gives the convergence of the formula for $n\to\infty$.

Generalization of the Trotter-Suzuki decomposition to the higher orders reads as
\begin{equation}
e^{x(A+B)}=e^{p_1xA}e^{p_2xB}e^{p_3xA}e^{p_4xB}\cdots e^{p_MxB}+O(x^{m+1}),
\end{equation}
where the selection of the parameters $\left\{p_1, p_2, \cdots, p_M \right\}$ is carried out in such a way that the correction term is of the order of $x^{m + 1}$. In this paper we will benchmark only first and second order TS decompositions that are more relevant from practical point of view.

\begin{figure}[t]
\includegraphics[width=0.95\columnwidth]{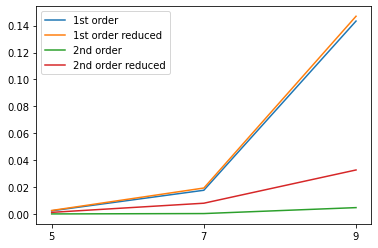}
\caption{Error of the TS decomposition versus number of qubits for the dense case. The Hamiltonian has 105 randomly sampled Hermitian terms and about 30 of them are weaken in 50 times. The sensitivity analysis allows to reduce number of exponentiations to 95, 76, 75 terms for 5, 7, 9 qubits respectively with moderate increasing of the error. }
\label{dense}
\end{figure}

\section{Variance-based global sensitivity analysis}

Modern physical systems can be very complex and when we try to model them the relationships between inputs and outputs are mostly poorly understood. The inputs and outputs can be subject to different sources of uncertainty, including external noise, errors in measurements and incomplete understanding of intrinsic mechanisms that manage the complex system under consideration. This uncertainty imposes a limit on our confidence in the output of the model and requires an knowledge of how much each input is contributing to the output uncertainty. Sensitivity analysis performs the role of ordering the inputs by impact in the variance of the output. In models involving many input variables, sensitivity analysis allows to make the dimension reduction. Here we overview the theory of variance-based global sensitivity analysis \cite{sobol1993,sobol2001,saltelli2010,saltelli2008}.

\subsection{Decomposition of variance and Sensitivity indexes}

From a black box perspective, any model may be viewed as a function $y=f(x)$ defined in the $d$-dimensional unit hypercube, where $x$ is a vector with $d$ components $x_1, x_2, \ldots x_d$, and $y$ is a scalar output. In order to intuitively understand how the sensitivity analysis works lets consider what happens with the output $y$ when we fix one of the variables $x_i$ at a particular value $\tilde{x_i}$. Denote the conditional variance of $y$ over all variables but $x_i$ as $Var_{x_{-i}}(y|x_i=\tilde{x_i})$. It is evident that the resulting conditional variance with one frozen variable is less or equal to the total unconditional variance $Var(y)$. Furthermore, it is easy to notice that the smaller $Var_{x_{-i}}(y|x_i=\tilde{x_i})$ in comparison with the total variance of the model, the greater the influence of $x_i$ on the model. Thus, the conditional variance $Var_{x_{-i}}(y|x_i=\tilde{x_i})$ can be used as a measure of the relative importance of $x_i$. In order to avoid the dependence of the measure on the particular $\tilde{x_i}$ we can average $Var_{x_{-i}}(y|x_i=\tilde{x_i})$ over all possible points $x_i$ in our hypercube and consider the expectation value $E_{x_i}(Var_{x_{-i}}(y|x_i))$ as the importance measure. This expectation value satisfies the following expression:
$$
E_{x_i}(Var_{x_{-i}}(y|x_i))+Var_{x_i}(E_{x_{-i}}(y|x_i)) = Var(y)
$$ 
and a small $E_{x_i}(Var_{x_{-i}}(y|x_i))$ (or large $Var_{x_i}(E_{x_{-i}}(y|x_i))$) implies that $x_i$ is an important variable in the model. Defining and ordering the sensitivity indexes of the first order as
$$
S_i = \frac{Var_{x_i}(E_{x_{-i}}(y|x_i))}{Var(y)}
$$ 
it is possible to estimate how much each input is contributing to the output uncertainty (a high value indicates an important variable).

More rigorously and generally speaking, according to the Sobol's approach the $y=f(x)$ may be decomposed as

\begin{equation}
\label{exp_decomp}
y = f_0 + \sum_{i=1}^d f_i(x_i) + \sum_{i<j}^d f_{ij}(x_i, x_j) + \ldots + f_{1,2,\ldots,d}(x_1,\ldots,x_d)
\end{equation}
with the conditions that $f_0$ is constant and the integrals 
\begin{equation}
\label{cond}
\int_0^1 f_{i_1,\ldots, i_s}(x_{i_1},\ldots,x_{i_s}) dx_{i_k} = 0, i_k\in\{i_1,\ldots, i_s\}.
\end{equation}

From the definition it follows that $f_0 = \int f(x) dx$ and the different terms in the decomposition are pairwise orthogonal due to (\ref{cond}) and in fact if $\{i_1,\ldots, i_s\}\neq \{j_1,\ldots, j_l\}$ then $\int f_{i_1,\ldots, i_s}(x_{i_1},\ldots,x_{i_s}) f_{j_1,\ldots, j_l}(x_{j_1},\ldots,x_{j_l}) dx =0$.


The process of proving (\ref{exp_decomp}) can be sketched as following (see \cite{sobol1993} for the details): 
\begin{itemize}
	\item Consider the set of functions $\{g_i, g_{ij}, g_{ijk}\ldots\}$ defined as integrals over all variables but $\{\{x_i\}, \{x_i,x_j\}, \{x_i,x_j,x_k\}\ldots\}$ (for example, $g_i = \int\ldots\int f(x) dx_{-i}$, $g_{i,j} = \int\ldots\int f(x) dx_{-ij}$ and so on, where $dx_{-i}$ denotes integration over all variables except the variable $x_i$ over which the function is not integrated);
	\item Integrate both sides of (\ref{exp_decomp}) over all variables except $\{x_i\}$, $\{x_i,x_j\}$, $\{x_i,x_j,x_k\} \ldots$ and taking into account the condition of (\ref{cond}) we will get $$g_i = f_0 + f_i(x_i),$$ $$g_{ij} = f_0 + f_i(x_i) + f_j(x_j) + f_{ij}(x_i,x_j),\ldots$$ and thus we can find all the terms in the decomposition of (\ref{exp_decomp}). 
\end{itemize}

Without any loss of generality, it is always possible to map the unit hypercube to a desired distribution with the joint probability distribution $p(x)$ and in this case the terms in decomposition can be rewritten as the expectation values in integral form

\begin{multline}
\label{terms_exp}
f_0 = E(y) =\int f(x)p(x)dx \\
= \int\ldots\int f(x_1,\ldots,x_d)p(x_1)\ldots p(x_d) dx_1\ldots dx_d,\\
f_i(x_i) = E(y|x_i) - f_0 = \int f(x)p(x_{-i})dx_{-i} -f_0,\\
f_{ij}(x_i, x_j) = E(y|x_i,x_j) - f_0 - f_i - f_j \\
= \int f(x)p(x_{-ij})dx_{-ij} - f_0 - f_i - f_j,\\
\ldots\\
f_{1,2,\ldots,d}(x_1,\ldots,x_d) = E(y|x_1,\ldots,x_d) \\- (f_0 + \sum_{i=1}^d f_i(x_i) + \sum_{i<j}^d f_{ij}(x_i)+\ldots),
\end{multline}
where $p(x_{-i})$ is a product of probability distributions for all variables but $x_i$.

Assuming that the $f(x)$ is square-integrable, the decomposition of (\ref{exp_decomp}) can be squared and integrated and we can rewrite the equation in terms of variances

\begin{equation}
\label{var_decomp}
Var(y) = \sum_{i=1}^d V_i + \sum_{i<j}^d V_{ij} + \ldots + V_{12\ldots d}
\end{equation}

where 

\begin{multline}
Var(y) =\int f^2(x) p(x) dx - f_0^2 =\int (f(x)-E(y))^2 p(x) dx, \\
V_i = Var_{x_i}(E_{x_{-i}}(y|x_i)) = \int f_i^2(x_i)dx_i,\\
V_{ij} = Var_{x_{ij}}(E_{x_{-ij}}(y|x_i,x_j))-V_i-V_j \\
= \int f_{ij}^2(x_i,x_j)dx_i dx_j,
\end{multline}

and so on. Here, $V_i$ defines the effect of varying $x_i$ alone (the first-order interaction effect), and $V_{ij}$ is the effect of varying $x_i$ and $x_j$ simultaneously (the second-order interaction effect) and so on ($x_{-i}$ denotes all variables except $x_i$). This representation shows that the variance of the model output can be decomposed into terms respective to each input and the interactions between them. All the terms sum up to the total variance of the model. Dividing the both parts of equation of (\ref{var_decomp}) by the total variance it is possible to define the sensitivity indexes of higher orders as following

\begin{equation}
\label{sens_indxs}
S_{\alpha\beta\ldots} = \frac{V_{\alpha\beta\ldots}}{Var(y)}
\end{equation}
and the equation reads as
\begin{equation}
\label{sens_decomp}
\sum_{i=1}^d S_i + \sum_{i<j}^d S_{ij} + \ldots + S_{12\ldots d} = 1
\end{equation}

These sensitivity indexes allow to identify the inputs that have no effect on the output and remove redundant parts of the model.

\subsection{Estimators for sensitivity indexes}

In the most cases it is not possible to calculate the sensitivity indexes analytically and usually they are estimated with some numerical approximation methods. One of the most common approaches in this case is the Monte Carlo method. The Monte Carlo method involves generating a sequence of randomly distributed points in the input space and calculating some probabilistic characteristics of the process under consideration. In practice, in order to improve efficiency, the quasi-Monte Carlo method is used, in this case the genuine random sample sequence replaced with low-discrepancy sequence. One of the low-discrepancy sequences commonly used in sensitivity analysis is the Sobol sequence. In order to calculate the sensitivity indexes in the Sobol paradigm, the following steps are required: 1) Generate an $N\times 2d$ sample matrix where each row is a sample point with $2d$ dimensions; 2) Denote the first $d$ columns of the matrix as matrix $A$, and the remaining $d$ columns as matrix $B$. This effectively can be considered as two independent sets of samples of N points in the $d$-dimensional unit hypercube; 3) Construct a set of $d$ matrices with dimension of $N\times d$ with replacing of $i$th column of matrix $A$ with $i$th column of matrix $B$ and the remaining columns leave unchanged, denote each such matrix as $A_i(B)$ where $i = 1,2,...,d$; 4) The matrices $A$, $B$, and $A_i(B)$ define $N\times (d+2)$ points in the input space (one for each row). Using this sample points calculate the corresponding outcome values $f(A)$, $f(B)$ and $f(A_i(B))$; 5) Then in order to calculate the sensitivity indexes it is possible to use the following estimators:

\begin{equation}
\label{estimators}
Var_{x_i}(E_{x_{-i}}(y|x_i)) \approx \frac1{N} \sum_{n=1}^{N} f(B)_n(f(A_i(B))_n - f(A)_n)
\end{equation}

It is worth to emphasize that the accuracy of the estimators depends on $N$ which can be chosen by sequentially adding points to reach convergence. In order to estimate the sensitivity indexes for all input variables it is required to run calculations $N(d+2)$ times. Since N is often quite large, computational cost can quickly become challenging when the model requires a significant amount of time for a single run. In order to reduce the computational cost some additional numerical techniques were developed, such as emulators, HDMR \cite{rabitz89,li06,li02} and FAST \cite{Cukier1973}.

\section{Optimization of the Trotter-Suzuki decomposition}

In this section we present an algorithm for optimization of the TS decomposition by removing unimportant gates from consideration using the global sensitivity analysis with some increasing in error. We apply the algorithm to the first order TS decomposition and show that it allows to find unimportant terms in the system Hamiltonian. We present a proof-of-concept examples with randomly sampled dense and sparse Hermitian matrices as parts of a system Hamiltonian. 

\begin{table}
\begin{center}
\begin{tabular}{ |c|c|c|c| } 
 \hline
 number of qubits & 5 & 7 & 9 \\ 
 \hline
 dense (no. of gates after reduction) & 95 & 76 & 75 \\ 
 \hline
 sparse (no. of gates after reduction) & 104 & 95 & 77 \\ 
 \hline
\end{tabular}
\caption{Number of gates after truncation for different number of qubits for dense and sparse cases. Initial randomly generated system Hamiltonians consist of 105 terms for each number of qubits.}
\label{table1}
\end{center}
\end{table}

In order to construct the proof-of-concept examples, we randomly generate Hamiltonians with some fixed number of random Hermitian matrices, sparse (only 30 percent of each Hamiltonian term are non-zero) and dense:
\begin{equation}
H = \sum_n H_n.
\end{equation}
Our algorithm is based on sensitivity analysis of the Frobenious norm of different linear compositions of the terms in our system Hamiltonian
\begin{equation}
\label{norm}
\left\|\sum_n \beta_{in} H_n\right\|
\end{equation}
where the coefficients $\beta_{in}$ constructed as quasi-random Saltelli sequence \cite{saltelli2008} of the coefficients and $i$ is indexing similar for rows in matrices $A$, $B$, and $A_i(B)$ from the previous section. Then we calculate how the coefficients affect the variance of the norm of the system Hamiltonian of (\ref{estimators}). In order to take into account contribution of each Hamiltonian term in the total output variance we consider sensitivity indexes of first-order only of (\ref{sens_indxs}) and trying to indicate with the indexes what terms of our Hamiltonian can be truncated. We use the SALib python library \cite{salib} to implement our calculations of the sensitivity indexes on the HPC \cite{hpc}.

\begin{figure}[t]
\includegraphics[width=0.95\columnwidth]{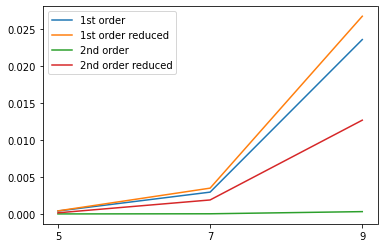}
\caption{Error of the TS decomposition versus number of qubits for the sparse case. The Hamiltonian has 105 randomly sampled Hermitian terms and about 30 of them are weaken in 50 times. The sensitivity analysis allows to reduce number of exponentiations to 104, 95, 77 terms for 5, 7, 9 qubits respectively with moderate increasing of the error. }
\label{sparse}
\end{figure}

In particular, we consider quantum systems with 5, 7 and 9 qubits. For each of them we construct a Hamiltonian on the basis randomly sampled Hermitian matrices with 105 terms $H= \sum_n H_n$ (separately for dense and sparse cases). Then we artificially weaken about 30 of randomly choosen $\{H_n\}$ in 50 times multiplying with 0.02 and construct a Hamiltonian $\tilde{H} = \sum_n \tilde{H}_n$ with the weaken terms and the rest (the total number of terms in the Hamiltonian $\tilde{H} $ does not change and still equals to $105$). In order to calculate error of the TS decomposition in comparison with genuine evolutionary operator with Hamiltonian $\tilde{H}$ we calculate the norm of difference of matrix exponentiations:
\begin{equation}
\label{error1}
\epsilon_1 = \left\|e^{-i\tilde{H}t}-\prod_{n=1}^{N} e^{-i\tilde{H}_n t}\right\|,
\end{equation}
for the first-order TS approximation and 
\begin{equation}
\label{error2}
\epsilon_2 = \left\|e^{-i\tilde{H}t}-\prod_{n=1}^{N} e^{\frac{-i\tilde{H}_n t}{2}}\prod_{l=N}^{1} e^{\frac{-i\tilde{H}_l t}{2}}\right\|,
\end{equation} 
for the second-order TS approximation. In order to calculate the errors we use the HiQPulse python library \cite{hiq_lib} which allows to calculate the matrix exponentiations more than 5 times faster in comparison with SciPy library.

Then we implement the sensitivity analysis for the Frobenious norm of $\left\|\beta_{i}\tilde{H}\right\| = \left\|\sum_n \beta_{in} \tilde{H}_n\right\|$ with the use of the Saltelli sequence and find the first-order sensitivity indexes. We remove from consideration the terms with sensitivity indexes which are in 1000 times smaller than the average value of the sensitivity indexes for the dense case and in 5000 times smaller than the average value for the sparse case. After that, we calculate the error of the TS decomposition for the reduced Hamiltonian in comparison with genuine evolutionary operator with Hamiltonian $\tilde{H}$ using the HiQPulse python library. It allows to remove 4, 13 and 25 gates in the first-order TS decomposition for 5, 7 and 9 qubit systems respectively for the dense case (6, 11 and 26 gates for the sparse case) with very small increasing in error (estimated analogously to (\ref{error1}) and (\ref{error2})), see Table \ref{table1}. 

Choosing the same terms for reduction in the second-order TS decomposition we can reduce number of gates from 209 in about 5, 13 and 25 percents for 5, 7 and 9 qubit systems respectively with moderate increasing in error, see Table \ref{table1}.   

On the Fig. \ref{dense}, the results in error between actual system evolution and TS approximations with and without removing unimportant gates in the TS approximations of different orders for the dense case are presented. We can see that the error of the reduced second-order TS approximation is much better in comparison with the first-order TS decomposition with 50 percent more gates for implementation and moderately worse than the genuine second-order TS approximation with 25 percent less number of gates. This opens up some opportunities for more optimal TS approximations with truncation unimportant gates in the NISQ era devices. The similar results are obtained for the sparse case (see Fig. \ref{sparse}).

\section{Discussion}

Thus we have shown that the Global sensitivity analysis can be used for removing from TS decomposition unimportant gates in quantative way and allows to design quantum circuit in more flexible way. The approach is based on removing the terms that give small contribution in the whole problem variance that can be indicated with the sensitivity indexes. Further development of the method can be around consideration of sensitivity indexes of higher orders that allows to take into account cross interaction in system Hamiltonian.


\begin{thebibliography}{999}
\bibitem{hiq_lib} Will be available on the GitHub later.
\end{thebibliography}
\end{document}